\documentclass[12pt,preprint]{aastex}

\usepackage{graphics,graphicx}
\usepackage{amsmath}
\usepackage{natbib}

\usepackage{color}

\def\apjl{\rm ApJL}
\def\apjs{\rm ApJS}

\def\aap{\rm AAP}

\newcommand{\gtorder}{\mathrel{\raise.3ex\hbox{$>$}\mkern-14mu
            \lower0.6ex\hbox{$\sim$}}}
\newcommand{\ltorder}{\mathrel{\raise.3ex\hbox{$<$}\mkern-14mu
            \lower0.6ex\hbox{$\sim$}}}

\shorttitle{Retrograde Binary Evolution}
\shortauthors{Schnittman \& Krolik}

\begin{document}

\title{Evolution of a Binary Black Hole with a Retrograde Circumbinary
  Accretion Disk} 

\author{Jeremy D. Schnittman}
\affil{NASA Goddard Space Flight Center, Greenbelt, MD 20771 }
\affil{Joint Space-Science Institute, College Park, MD 20742 }
\author{Julian H. Krolik}
\affil{Department of Physics and Astronomy, Johns Hopkins University,
  3400 N Charles Street, Baltimore, MD 21218} 

\begin{abstract}
We consider the evolution of a supermassive black hole binary (SMBHB)
surrounded by a retrograde accretion disk.  Assuming the disk is exactly
in the binary plane and transfers energy and angular momentum to the binary via
direct gas accretion, we
calculate the time evolution of the binary's semi-major axis $a$ and
eccentricity $e$.  Because the gas is predominantly transferred when the
binary is at apocenter, we find the eccentricity grows rapidly while
maintaining constant $a(1+e)$.   After accreting only a fraction of the
secondary's mass, the eccentricity grows to nearly unity; from then on,
gravitational wave emission dominates the evolution, preserving constant
$a(1-e)$.  The high-eccentricity waveforms redistribute the peak
gravitational wave power from the nHz to $\mu$Hz bands, substantially
affecting the signal that might be detected with pulsar timing arrays. 
We also estimate the torque coupling binaries of arbitrary eccentricity
with obliquely aligned circumbinary disks.   If the outer edge of the disk
is not an extremely large multiple of the binary separation, retrograde
accretion can drive the binary into the gravitational wave-dominated state
before these torques align the binary with the angular momentum of the
mass supply.
\end{abstract}

\section{Introduction}

The orbital evolution of binary black holes has been a subject of
considerable interest for more than thirty years, dating
back to the work of \cite{Begelman1980}. It has only grown in
prominence in recent years, as it has become apparent both that most
reasonably-large galaxies contain supermassive black holes in their
nuclei \citep{Kormendy1995,Magorrian1998} and that today's galaxies are the product of numerous
mergers between smaller galaxies \citep{Bell2006,McIntosh2008,deRavel2009,Bridge2010}, each of which likely
contains its own nuclear black hole.   As pointed out by
\cite{Begelman1980}, when two massive black holes in a galaxy approach
near enough one another to form a gravitationally-bound binary,
interactions with external stars become much less effective at
bringing them still closer together. A number of processes have since 
been suggested as viable mechanisms to push such a system closer
toward the merger of the two black holes themselves \citep{Merritt2005}. 

A particularly popular such mechanism
\citep{GouldRix2000,Armitage2002,Escala2004,Escala2005,MM2008,Cuadra2009} 
is accretion of gas onto the binary black hole through a circumbinary
disk.   In principle, the orbital plane of such a disk could be wholly
independent of the orbital plane of the binary, particularly at large
radius. However, a number of processes have been pointed 
out that tend to push the two orbital planes toward coincidence
\citep{Nixon2012,MK2013}.   Whether the angular momenta of the two
orbits align or counter-align, however, is less clear. 

Retrograde circumbinary disks have certain special properties
\citep{Nixon2011,Roedig2014,Bankert2014}. Because the angular momentum
acquired by accretion is directed opposite to the binary orbital
angular momentum, accretion in such a context tends to drive a rapid 
decrease in the orbital angular momentum while only modestly
decreasing the orbital energy.    The result is an
increase in the orbital eccentricity.  Indeed, the first paper to
consider evolution of this sort \citep{Nixon2011} concluded on the
basis of qualitative arguments that this eccentricity growth causes
the pericenter distance $r_p=a(1-e)$ to become so small that
gravitational wave (GW) radiation rapidly drives the system to merger.
Although GW radiation generically {\it decreases} eccentricity
\citep{Peters1964}, in their view the binary orbit would retain non-zero
eccentricity all the way to merger.

Here we study this problem quantitatively.
For a wide range of reasonable initial conditions and physical
assumptions, we find the same robust result: rapid growth in
eccentricity with constant apocenter $a(1+e)$ over a time long enough to
accrete a mass comparable to that of the secondary (for equal-mass
systems accreting at the Eddington limit this is a Salpeter time,
$t_{\rm Sal} = 4.5 \times 10^7$ years, for a nominal radiative
efficiency $\eta=0.1$). 
At that point the eccentricity reaches $e\gtrsim 0.99$, and
gravitational wave evolution takes over. In this phase, GW losses
occur almost entirely at pericenter, leading to evolution with
constant $a(1-e)$ until the orbit is nearly circular at a binary
separation of $\sim 100r_g$ (the gravitational radius $r_g \equiv GM/c^2$).

This evolutionary behavior can have profound effects on pulsar timing
array (PTA) observations of gravitational waves radiated by supermassive
black hole mergers.   If gas inflow in the center of a post-merger
galaxy takes place in a stochastic manner, retrograde disks should be roughly
as common as prograde disks, but much more efficient at effecting the binary
merger.   Rapid evolution from initial separations of $a \sim 10^5
r_g$ down to $a \sim 10^2 r_g$ will greatly reduce the number of
systems potentially observable in this regime \citep{Kocsis2011,Sesana2013,Ravi2014}.
As a result, the strength \citep{Nanograv2013} of the stochastic GW signal in the
PTA band $f \sim 10^{-9}-10^{-8}$~Hz would be sharply diminished.   In addition, the
early onset of GW-dominated evolution could lead to a decoupling
of the circumbinary disk from the inner binary at much greater radii than is typically assumed,
altering electromagnetic signals associated with accretion onto the black
holes \citep{Milos2005,Chang2010,Tanaka2010,Noble2012,Schnittman2013,Schnittman2014,Farris2014}.

\section{Accretion-driven Evolution}\label{accretion_evolution}

As shown in a companion paper describing magneto-hydrodynamic
simulations of a retrograde circumbinary accretion disk
\citep{Bankert2014}, the nearly axisymmetric density profile of the
disk results in a very small gravitational torque on the binary. Thus,
unlike the prograde case, the primary means of energy and angular
momentum transfer from the disk to the binary is through direct
accretion. 

The eccentricity $e$ and semimajor axis $a$ of the binary can be written as
\begin{subequations}
\begin{equation}
  e = \left[1+\frac{2EJ^2(1+q)^6}{G^2M^5q^3}\right]^{1/2}
\end{equation}
and
\begin{equation}
  a = -\frac{q}{(1+q)^2}\frac{GM^2}{2E}\, ,
\end{equation}
\end{subequations}
with $M$ the binary's total mass, $J$ angular momentum, $E$ energy,
and mass ratio $q\equiv M_2/M_1\ le 1$. The evolution equations are
then given by 
\begin{subequations}
\begin{equation}\label{eqn:orbevol}
\dot e = \frac{1 - e^2}{2e} \left[ -2 \dot J/J - \dot E/E +
  5\dot M/M + 3 \frac{1-q}{1+q} \dot q/q \right]
\end{equation}
and
\begin{equation}\label{eqn:orbevol2}
\dot a/a = -\dot E/E + 2\dot M/M + \frac{1-q}{1+q} \dot q/q. 
\end{equation}
\end{subequations}

All of the terms on the right hand sides of both equations can be
written in terms of the fractional accretion rate $\dot{M}/M$. If the specific angular
momentum of the accreting material is $j_{\rm gas}$, then $\dot J/J =
[j_{\rm gas}/(J/M)] \dot{M}/M$. We find it convenient to parameterize
$j_{\rm gas}$ in terms of the canonical Keplerian momentum via $j_{\rm
  gas}= \mathcal{J}\sqrt{GMa}$, giving 
\begin{equation}\label{eqn:Jdot_J}
\frac{\dot{J}}{J} = \mathcal{J}\frac{(1+q)^2}{q}
\frac{1}{(1-e^2)^{1/2}}\frac{\dot{M}}{M}\, .
\end{equation}
>From the MHD simulations of an equal-mass, circular retrograde binary in
\citet{Bankert2014}, we find $\mathcal{J}\approx -1.25$. Assuming the
inner edge of the circumbinary disk is limited by the radius of the
binary orbit at apocenter, we believe it is reasonable to scale this term
with eccentricity as $\mathcal{J}\approx -1.25\sqrt{1+e}$. 

We parameterize the
relative fraction of gas accreting onto each black hole as $f_1$ and
$f_2$ ($f_1 + f_2 = 1$), giving an expression for $\dot{q}$:
\begin{equation}
\frac{\dot q}{q} = \frac{1+q}{q}(f_2 - qf_1)\frac{\dot M}{M}\, .
\end{equation}
Here and in the following, quantities subscripted by 1 pertain to the primary,
those subscripted by 2 pertain to the secondary.

The relation between $\dot{E}$ and $\dot M$ is harder to determine
because it depends on how much dissipation is associated with
attaching the orbital streams to the members of the binary.   Because
shocks are extremely likely, dissipation will in most instances be
very significant.  Here we choose to estimate the energy delivered by
accretion on the basis that the system is maximally dissipative, that is,
that shocks are capable of dissipating as much orbital energy into heat
(and likely radiating it) as is consistent
with momentum conservation. In other words, the gas streams inward
from the circumbinary disk and then collides inelastically with one or
the other of the ``minidisks", the individual accretion disks around
each black hole.

We assume that the accretion stream encounters the minidisk when the
binary is at apocenter. Although we cannot confirm this assumption
with our simulation data (the binary in the simulation had zero
eccentricity), it is quite plausible: after all, this is the orbital 
phase at which the members of the binary are closest to the
circumbinary disk, and also the phase at which the binary spends most
of its time.   

At apocenter, the total energy in the orbit can be written as
\begin{eqnarray}
E &=& -\frac{GM_1 M_2}{2a}= -\frac{GM^2}{2a}\frac{q}{(1+q)^2}
\nonumber\\
&=& -\frac{GM^2}{a(1+e)}\frac{q}{(1+q)^2}+\frac{1}{2}M_1 v_1^2
+\frac{1}{2}M_2 v_2^2\, ,
\end{eqnarray}
where in the second line we separate the energy into potential and
kinetic terms, with $v_{1,2}$ the velocities at apocenter, and
$r_{1,2} = a(1+e)(q,1)/(1+q)$ the radial positions of the two black holes.
Here we adopt a notation where the respective quantities for the two black holes are
written $(1,2)$. The rate of change in the binary's potential energy is then
\begin{eqnarray}
\dot{E}_p = -\frac{G(M_2\dot{M_1}+\dot{M_2}{M_1})}{a(1+e)} &=& 
-\frac{GM^2}{a(1+e)}\frac{\dot{M}}{M}\left(\frac{q}{1+q}f_1+\frac{1}{1+q}f_2\right)
\nonumber\\
&=& 2E\frac{\dot{M}}{M}\frac{1}{1+e}\frac{1+q}{q}(qf_1+f_2)\, .
\end{eqnarray}

For the kinetic energy terms, it is convenient to write the kinetic
energy at apocenter in terms of the linear momentum of each BH:
$E_{k1,2} = 1/2(p_{1,2}^2/M_{1,2})$. Then we have
\begin{equation}
\dot{E}_{k1,2} = \frac{p_{1,2}\dot{p}_{1,2}}{M_{1,2}}-
\frac{p^2_{1,2}}{2M^2_{1,2}}\dot{M} f_{1,2} = 
v_{1,2}\dot{p}_{1,2}-\frac{1}{2}v^2_{1,2}\dot{M}f_{1,2}
\end{equation}
where the linear velocities at apocenter are
\begin{equation}
v_{1,2} = \sqrt{\frac{GM}{a}}\sqrt{\frac{1-e}{1+e}}\frac{(q,1)}{1+q} 
\equiv v_{\rm apo}\frac{(q,1)}{1+q}\, .
\end{equation}

The momentum change comes from the accretion stream:
\begin{equation}
\dot{p}_{1,2} = v_{{\rm gas}1,2}\dot{M} f_{1,2}
\end{equation}
which has velocity 
\begin{equation}
v_{{\rm gas}1,2} = \frac{j_{\rm gas}}{r_{1,2}}\, .
\end{equation}
As above, $j_{\rm gas}$ is the specific angular momentum of the gas in the
accretion stream. We can now write
\begin{equation}
v_{{\rm gas}1,2} =
\frac{\mathcal{J}\sqrt{GMa}}{a(1+e)}\frac{1+q}{(q,1)} =
\mathcal{J}\sqrt{\frac{GM}{a}}\sqrt{\frac{1-e}{1+e}}\sqrt{\frac{1}{1-e^2}}
\frac{1+q}{(q,1)} = \mathcal{J}v_{\rm
  apo}\sqrt{\frac{1}{1-e^2}}\frac{1+q}{(q,1)}\, .
\end{equation}

Now we can find the change in kinetic energy:
\begin{eqnarray}
\dot{E}_{k1,2} &=& v_{\rm
  apo}^2\sqrt{\frac{1}{1-e^2}}\mathcal{J}\dot{M}f_{1,2}
-\frac{1}{2}v_{\rm apo}^2\frac{q^2,1}{(1+q)^2}\dot{M}f_{1,2}
  \nonumber\\
&=& -\dot{M}f_{1,2}v_{\rm apo}^2
\left[-\mathcal{J}\sqrt{\frac{1}{1-e^2}}+\frac{1}{2}\frac{(q^2,1)}{(1+q)^2}\right]\, .
\end{eqnarray}
Writing 
\begin{equation}
v_{\rm apo}^2 =
\left(\frac{-2E}{M}\right)\frac{1-e}{1+e}\frac{(1+q)^2}{q},
\end{equation}
and combining this with the expression for the potential energy, we get
\begin{equation}
\frac{\dot{E}_{1,2}}{E} = \frac{\dot{M}}{M}f_{1,2}\frac{1}{1+e}
\left[-2\mathcal{J}\sqrt{\frac{1-e}{1+e}}\frac{(1+q)^2}{q}+(1-e)(q,q^{-1})
+2\frac{1+q}{q}(q,1)\right]
\end{equation}
Recall that, since $\mathcal{J}<0$, this expression is always strictly
positive.   Therefore the binary always evolves toward greater binding energy.

We can now inspect equations (\ref{eqn:orbevol},\ref{eqn:orbevol2}) in
a few limiting cases: $q=1$, $q\ll 1$, $e\ll 1$, $1-e=\delta\ll
1$. The leading-order terms in each case are shown in Table
\ref{table:limits}. For $q\ll 1$ we assume $f_2=1$. Not surprisingly,
the eccentricity always grows and the semi-major axis always shrinks
for binaries in the presence of retrograde accretion disks. Also not
surprisingly, the effect is fastest when $q$ is small. Note that all
evolution terms are proportional to $q^{-1}\, \dot{M}/M$, so the characteristic
time scales can be rather shorter than the accretion time $M/\dot{M} =
 4.5\times 10^7 \dot{m}^{-1}$~yr, with $\dot{m}$ the Eddington-scaled
accretion rate for a radiative efficiency of $\eta=0.1$.

\begin{table}[ht]
\caption{\label{table:limits} Leading-order terms for $\dot{e}$ and
  $\dot{a}$ for limiting values of $q$ and $e$.}
\begin{center}
\begin{tabular}{cccc}
$q$ & $e$ & $\dot{e}$ & $\dot{a}/a$ \\
\hline
1       & $\ll 1$    & $13 \frac{\dot{M}}{M}$ & $-13 \frac{\dot{M}}{M}$ \\
$\ll 1$ & $\ll 1$    & $\frac{9}{2q} \frac{\dot{M}}{M}$ & $-\frac{9}{2q} \frac{\dot{M}}{M}$ \\
1       & $1-\delta$ & $10\sqrt{\delta} \frac{\dot{M}}{M}$ &
$-5\sqrt{\delta} \frac{\dot{M}}{M}$ \\
$\ll 1$ & $1-\delta$ & $\frac{2.5}{q}\sqrt{\delta}
\frac{\dot{M}}{M}$ & 
$-\frac{1.25}{q}\sqrt{\delta} \frac{\dot{M}}{M}$ \\
\end{tabular}
\end{center}
\end{table}

Because we assume that all accretion takes place at apocenter, and the
accretion flow is tangent to the binary orbit at that point, the
impulse approximation implies that $a(1+e)$ should remain
constant. When $e\ll 1$, this implies that $\dot{e}=-\dot{a}/a$, while
$\dot{e}=-2\dot{a}/a$ when $e\approx 1$. This results are consistent with
the limiting behavior shown in Table \ref{table:limits}. 

\section{Gravitational Wave-Driven Evolution}

In the previous section, we saw that the binary eccentricity can grow
rapidly via retrograde accretion. \citet{Nixon2011} also reached this conclusion,
and predicted that, as eccentricity grew and the pericenter separation
decreased, gravitational waves would take over as the driving
evolutionary mechanism of the system. Here we present a more detailed
analysis of this process. \citet{Peters1964} gives the
gravitational-wave-driven evolution of eccentricity and separation in
the post-Newtonian regime:
\begin{subequations}
\begin{eqnarray}\label{eqn:adot_gw}
(\dot a/a)_{\rm GW} &=& -\frac{64}{5}\frac{q}{(1+q)^2} (1 -
  e^2)^{-7/2}\left(1+\frac{73}{24}e^2+\frac{37}{96}e^4\right)
\left(\frac{r_g}{a}\right)^4 t_g^{-1}\\
\label{eqn:edot_gw} 
(\dot e)_{\rm GW} &=& -\frac{304}{15} \frac{q}{(1+q)^2} (1 -
  e^2)^{-5/2} 
e\left(1+\frac{121}{304}e^2\right)\left(\frac{r_g}{a}\right)^4 t_g^{-1}\, .
\end{eqnarray}
\end{subequations}
We have scaled the spatial dimensions by the gravitational radius $r_g
\equiv GM/c^2$ and the time scale by $t_g = r_g/c$. 

\begin{figure}[ht]
\begin{center}
\includegraphics[width=0.8\textwidth]{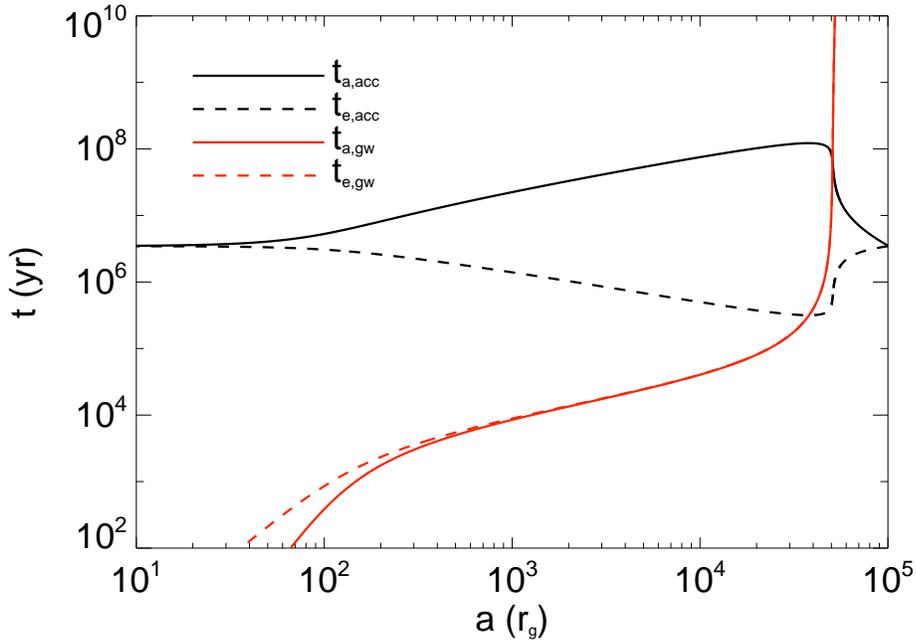}
\caption{\label{fig:timescales} Characteristic time scales for various
  evolutionary processes. As the system evolves from right to left,
  first the eccentricity rises rapidly with constant $a(1+e)$ and then
  gravitational-wave-driven evolution takes over, decreasing both $a$
  and $e$, while keeping $a(1-e)$ roughly constant. We take for
  initial values $M=10^8 M_\odot$, $q=1$, $\dot{m}=1$, $a_0=10^5 r_g$,
  and $e_0=0$.  $t_{\rm e,gw} = t_{\rm a,gw}$ over a broad range
  of semi-major axes because $a(1-e)$ is constant. }
\end{center}
\end{figure}

For a given $a$, $e$ we can now define the characteristic time scales
$t_e = (1-e)/\dot{e}$ and $t_a=a/\dot{a}$ 
for the competing evolutionary processes: accretion and gravitational
waves. For a fiducial binary system with initial parameters
$M=10^8M_\odot$, $q=1$, $\dot{m}=1$, $a=10^5 r_g$, and $e=0$, we plot
in Figure \ref{fig:timescales} the relevant timescales as a function
of $a(t)$. The graph should be read from right to left, as $a$ shrinks
with time. Whichever timescale is shortest at a given orbital
separation is the dominant process governing the binary evolution at
that point. 

Consider, for example, the evolution of our fiducial case.   As it accretes
from its circumbinary disk, its eccentricity grows
rapidly for a few million years, while its semi-major axis simultaneously shrinks
by a factor of two. At this point, the high eccentricity leads to an extremely
sharp decrease in $t_{a,{\rm gw}}$ and the system transitions from
accretion-driven to gravitational wave-driven evolution. For highly
eccentric orbits, most of the gravitational radiation is emitted at
pericenter passage, when the two bodies have the highest
velocity. This fact justifies an impulse approximation in which the
energy and angular momentum are lost instantaneously at pericenter, so
the binary evolves with constant $a(1-e)$. 

\begin{figure}[ht]
\begin{center}
\includegraphics[width=0.8\textwidth]{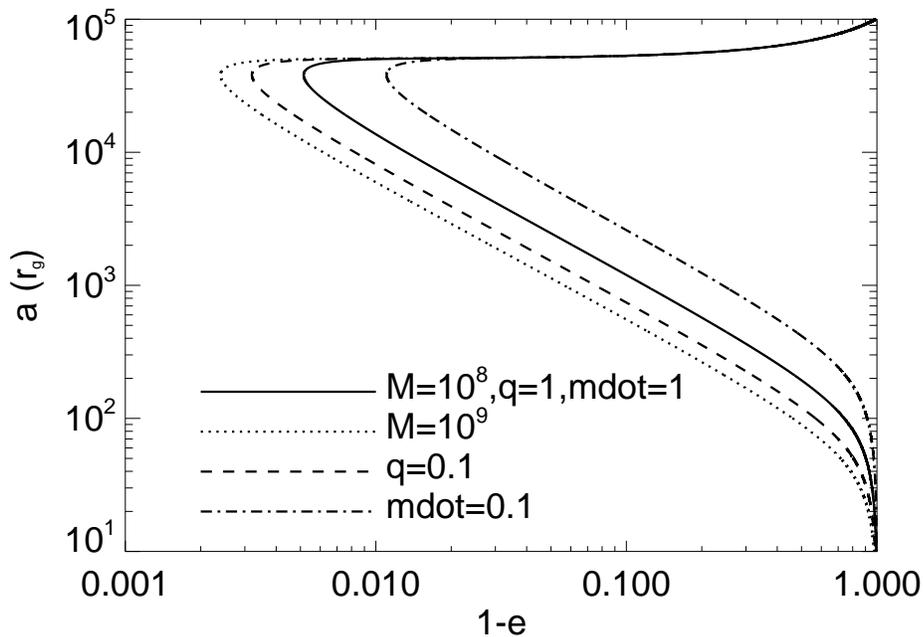}
\caption{\label{fig:a_vs_e} Evolution of $a$ and $e$ for the fiducial
  binary model with $M=10^8 M_\odot$, $q=1$, and $\dot{m}=1$ (solid
  curve). We also show variations from this model with $M=10^9
  M_\odot$ (dotted curve), $q=0.1$ (dashed curve), and $\dot{m}=0.1$
  (dot-dashed curve).}
\end{center}
\end{figure}

In Figure \ref{fig:a_vs_e} we show the mutual evolution of $a$ and $e$
for a few 
variations of the fiducial model parameters. In all cases the binary
begins in the upper-right corner with $a=10^5 r_g$ and $e=0$. The
dependence on the various parameters is straight-forward.  Because $t_g$
grows with $M$, whereas for fixed $\dot{m}$ the accretion time is
mass-independent, more
massive systems reach a higher eccentricity before gravitational
waves begin to dominate. For decreasing $\dot{m}$, the opposite
happens: the accretion time increases even though the gravitational time
remains constant, so the system leaves the accretion-dominated regime
earlier, at a smaller eccentricity. For small $q$, the gravitational
wave timescales increase and the accretion
timescales decrease; both effects combine to push the
system to higher eccentricity before entering the gravitational wave
regime.    More generally, because $a(1+e)$ is constant so long as
the evolution is accretion-dominated, and the transition to GW-driven
evolution occurs when $e \simeq 1$, the semimajor axis at this transition
is always $\simeq 1/2$ the initial semimajor axis. 

We can solve for the peak eccentricity $e_\ast$ by simply solving 
$\dot{e}_{\rm acc}=-\dot{e}_{\rm GW}$ in the limit of 
$1-e_\ast=\delta_\ast \ll 1$. For $\dot{e}_{\rm acc}$ we use equations
(\ref{eqn:orbevol}) and (\ref{eqn:Jdot_J}) and the observation that the
$\dot{J}/J$ terms dominate the evolution at high eccentricity. In this
limit we find
\begin{equation}\label{eqn:delta_trans}
\delta_\ast = 4 \times 10^{-3} \left[\frac{q^2}{(1+q)^4} a_5^{-4}
  M_8^{-1} \dot{m}^{-1}\right]^{1/3}\, ,
\end{equation}
which agrees quite well with the numerical evolutions plotted in
Figure \ref{fig:a_vs_e}. Here we use the scaling expressions
$a_5=a/(10^5 r_g)$ and $M_8 = M/(10^8 M_\odot)$. For a wide range of
parameters we find the time $t_{\rm trans}$ needed to evolve from
$e_0$ to $e_\ast$ can be approximated as
\begin{equation}\label{eqn:t_trans}
t_{\rm trans} \approx \frac{1}{3} \frac{q}{(1+q)^2}(1-e_0^2)^{1/2}
\dot{m}^{-1} t_{\rm Sal}\, .
\end{equation}
Because $q/(1+q)^2$, also called ``the symmetric mass ratio", is always
$\leq 1/4$, $t_{\rm trans} \le (1/12) (t_{\rm Sal}/\dot m)$, where
$(t_{\rm Sal}/\dot m)$ is the time over which
accretion increases the mass of the system by one $e$-fold.

\section{Gravitational Wave Signal}

For binaries on circular orbits, the gravitational waveform is a
sinusoidal signal with frequency exactly twice that of the orbital
frequency $f_{\rm orb}$.  For non-zero eccentricities, the
waveform is decidedly non-sinusoidal,
as the instantaneous orbital frequency increases as the black holes
move from apocenter to pericenter, peaking at 
\begin{equation}\label{eqn:f_p}
f_p = f_{\rm orb} \frac{(1+e)^{1/2}}{(1-e)^{3/2}}\, .
\end{equation}
Consequently, the range of harmonics carrying substantial power
increases with increasing eccentricity \citep{Tessmer2007,Enoki2007}.  

We calculate the amplitude $h(t)$ of the leading-order GW waveform from the time-varying
quadrupole moment: $h(t) \sim \ddot{Q}$. The power emitted in the
waveform is $L_{\rm GW}(t) \sim \dot{h}^2(t)$. Taking the Fourier
transform $L_{\rm GW}(f)$, we get the distribution of power in the individual
harmonics. To get the correct normalization for $L_{\rm GW}(f)$, we use the
orbit-averaged energy radiated $dE/dt$
\begin{equation}
\frac{dE}{dt} = \frac{GM^2 q}{(1+q)^2}\frac{\dot{a}}{2a^2}\, ,
\end{equation}
where $\dot{a}/a$ is given by \citet{Peters1964}. We find that, for
$1-e\ll 1$, the instantaneous $dE/df$ peaks around $f_p$. For
circular orbits $f_{\rm GW}=2f_p$, so the fact that $f_{\rm GW}
\approx f_p$ for eccentric orbits reflects the fact that the GW
radiation is emitted over a finite part of the orbit around
pericenter, during which $f<f_p$. 

\begin{figure}[ht]
\begin{center}
\includegraphics[width=0.8\textwidth]{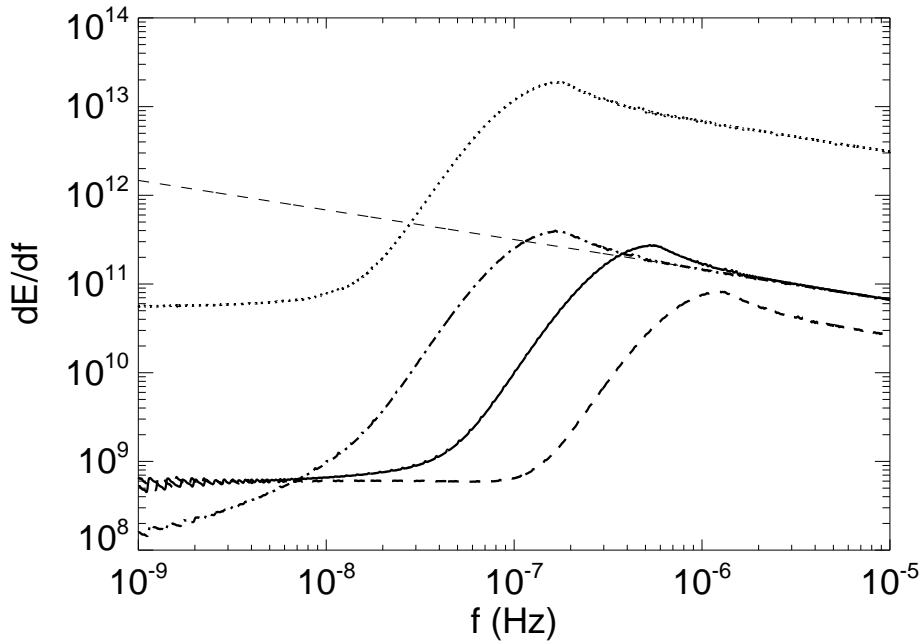}
\caption{\label{fig:dEdf} Gravitational-wave power spectrum for the
  fiducial binary model with $M=10^8 M_\odot$, $q=1$, and $\dot{m}=1$ (solid
  curve). We also show variations from this model with $M=10^9
  M_\odot$ (dotted curve), $q=0.1$ (dashed curve), and $\dot{m}=0.1$
  (dot-dashed curve). The thin dashed line is $dE/df\sim f^{-1/3}$.}
\end{center}
\end{figure}

Figure~\ref{fig:dEdf} shows the GW power spectrum for
the same four cases considered in Figure \ref{fig:a_vs_e}: fiducial
model with $M=10^8 M_\odot$, $q=1$, and $\dot{m}=1$ (solid
curve), along with variations from this model with $M=10^9
M_\odot$ (dotted curve), $q=0.1$ (dashed curve), and $\dot{m}=0.1$
(dot-dashed curve). The thin dashed line corresponds to $dE/df \sim
f^{-1/3}$, the classical result for purely circular evolution.

Figure~\ref{fig:dEdf} can best be understood by close comparison with Figure~\ref{fig:a_vs_e}.
As the binary's eccentricity increases through
accretion, the GW signal rises sharply in amplitude and frequency
until the binary evolution switches over to being dominated by
gravitational waves.  At this point, the binary evolves with constant $a(1-e)$, and thus
constant $f_{\rm GW}$, building up power in a single
peak. Eventually, the eccentricity is small enough that the
system follows the standard circular evolution track, with
both amplitude and frequency evolving to give $dE/df\sim
f^{-1/3}$.

Comparing it with Figure~\ref{fig:a_vs_e}, we can also easily
understand the relative offsets between the different models plotted
in Figure~\ref{fig:dEdf}.  For the more massive system with $M=10^9
M_\odot$ (dotted curve), the peak frequency shifts to the left since
the orbital frequency decreases with increasing mass. At the same
time, the GW amplitude increases because $dE/df \propto M^{5/3}$. For smaller
accretion rate ($\dot{m}=0.1$; dot-dashed curve), the peak
eccentricity is smaller, giving a lower peak frequency [see  eqn.\ (\ref{eqn:f_p})], but it
eventually evolves along exactly the same curve as the fiducial case. For
smaller mass ratio ($q=0.1$; dashed curve), the peak eccentricity
is larger, leading to a higher peak frequency. The amplitude scales
like $dE/df \propto q$, so the net effect is to shift the fiducial curve down and to
the right. 

The area under the solid black curve is actually the same as
that under the thin dashed line because they both represent the total
energy radiated going from $a\sim 10^5 r_g$ down to merger. Thus, all
the power normally radiated at low frequencies is concentrated in a
small peak around $f\sim 10^{-7}-10^{-6}$ Hz. This modification of the GW
spectrum can potentially have a major impact on the stochastic gravitational wave signal being
searched for by pulsar timing arrays (PTAs), which
are most sensitive to GW signals in the $10^{-9}-10^{-8}$ Hz range
\citep{IPTA2010,PPTA2013,Nanograv2013,Sesana2013}.  

If we take the initial separation to be somewhat smaller ($a=10^4 r_g$),
we find qualitatively similar behavior in the evolution, but from
equation (\ref{eqn:delta_trans}) we see the peak
eccentricity is ``only'' $e\approx 0.9-0.95$. In this case, the peaks
in Figure \ref{fig:dEdf} all shift to lower frequencies by a factor of
$\sim 2$. Integrating over the cosmic population of SMBHBs, this shift could
give a PTA stochastic signal with a peak around $f\approx
10^{-7}$ Hz and an amplitude roughly $50\%$ higher than the canonical
value, while sharply reducing the signal at lower frequencies.  

\section{Rapid Approach to Decoupling}

During the accretion-driven phase, the
binary separation $a(1+e)$ remains constant, so the radius at which
the secondary is farthest from the center-of-mass, and therefore best
able to capture accreting gas, remains unchanged.   Numerous studies
\citep{Roedig2014,DOrazio2013,Farris2014,Shi2014} have found that
so long as $q$ is not too small, the secondary captures the majority of
the accretion flow from a surrounding disk in prograde binaries; that fact
is likely to carry over to the retrograde case.    On the other hand, during
the GW-driven phase, $r_p=a(1-e)$ is fixed, while the apocenter
evolves like $r_a=r_p(1+e)/(1-e)$, which falls from $\simeq 2a_\ast$ to
$\simeq a_\ast(1-e_\ast)$ as the orbit grows less eccentric ($a_\ast$
and $e_\ast$ are the semi-major axis and eccentricity at the point
when the GW-driven phase begins).

The rapid diminution in apocenter distance during the GW-driven phase
may be faster than the characteristic inflow rate through the inner region
of the accretion disk.    If during the relevant phases of orbital evolution,
the disk is in the Shakura-Sunyaev ``outer zone'' (gas pressure-dominated,
free-free opacity greater than electron scattering opacity), this
inflow time at radius $r=r_5/(10^5 r_g)$ is
\begin{equation}\label{eqn:t_inflow}
t_{\rm inflow} \approx 7 \times 10^8 \, \dot{m}^{-3/10} \alpha_{-1}^{-4/5}
M_8^{6/5} r_5^{7/5} \hbox{~yr}\, .
\end{equation}
Here $\alpha=10^{-1}\alpha_{-1}$ is the parameter defined by
\cite{SS73} to describe the ratio of the vertically-integrated
internal stress to the vertically-integrated pressure, scaled to its
canonical value of $0.1$. While $t_{\rm inflow}$
is somewhat larger than both $t_{a,{\rm acc}}$ and $t_{e,{\rm acc}}$
during the accretion-dominated phase (see Fig.\ \ref{fig:timescales}),
this does not actually imply decoupling, since the important quantity
for comparison is the {\it apocenter} evolution timescale, which is
essentially infinite at this phase in the evolution, so the disk is
likely in inflow equilibrium. 

Equation (\ref{eqn:t_inflow}) assumes a steady-state thin disk
geometry, in which the local temperature is controlled by a balance between
heating due to accretion and local radiation.   However, it has been recognized for many years that
for typical AGN parameters, such disks are unstable to self-gravity at $r \gtrsim 10^3 r_g$
\citep{ShlosBegel1987,Goodman2003,JiangGoodman2011,Dunhill2014}, 
and this range of radii is of particular interest to the questions
studied here.   Unfortunately, little is known about what actually
happens as a result. It is possible, as these papers have discussed,
that the disk fragments and a significant part of its mass is
concentrated into new stars. 

On the other hand, we have observational evidence at larger scales
($\gtrsim 10^5 r_g$) that accreting material forms geometrically thick
structures, so-called ``obscuring tori;'' most recent work has supposed 
that they are supported by the AGN's radiation pressure interacting with dust grains in the gas
\citep{KP92,Chang2007,K2007,Dorodnitsyn2011,Dorodnitsyn2012,Roth2012}.  To account for
additional mechanisms that might alter the disk's vertical support, it is useful to
rephrase the inflow time in terms of the disk aspect ratio $h/r$, its ratio of vertical scale height
to radial position.   The inflow time then becomes
\begin{equation}
t_{\rm inflow} = 5 \times 10^7\, \alpha_{-1}^{-1} M_8 r_5^{3/2} 
\left(\frac{h/r}{0.01}\right)^{-2}\hbox{~yr}.
\end{equation}
We scale $h/r$ in this instance to $10^{-2}$ because the aspect ratio
predicted by classical disk theory is $2 \times 10^{-3} \alpha^{-1/10}
\dot m^{3/20} M_8^{-1/4} r_5^{1/20}$, and this is presumably a
lower bound.  

By comparison, the characteristic gravitational wave timescale $t_{\rm GW}=a/\dot{a}$
in the limit of $e\to 1$ is
\begin{equation}
t_{\rm GW} = 5\times 10^5\, \frac{(1+q)^2}{q} M_8\, a_5^4\,
\left(\frac{\delta_\ast}{0.003}\right)^{7/2}\hbox{~yr} .
\end{equation}
We have scaled to the typical orbital elements when gravitational
wave evolution begins to dominate.  We therefore expect that in most instances,
the apocenter changes too rapidly for the bulk of the disk to follow
once GW emission controls the system's evolution.

\citet{Milos2005} argued that whenever the binary evolution does become
faster than the disk's inflow time, mass inflow is cut off,
``decoupling" the disk from the binary and ending electromagnetic
emission due to accretion.  On the other hand, the global MHD
simulation of \cite{Noble2012} showed that the accretion rate can, in
fact, be sustained even in this regime---edge effects can
substantially raise the ratio of stress to pressure above its usual
value.   Similar effects are generically seen during the transient
phases of almost all global MHD accretion simulations
\citep{HGK2011,Sorathia2012}. In fact, analogous edge effects occur
even in purely hydrodynamic treatments \citep{Farris2014b}.   Thus,
the degree to which mass accretion actually diminishes when the
orbital evolution time is shorter than the disk inflow time is
uncertain. In either case, retrograde disks should reach this
``decoupling" considerably sooner than prograde disks.

\section{Disk-Binary Alignment}

A binary and its disk can remain in a stable counter-aligned
orientation if $J_{\rm disk}<2J_{\rm bin}$
\citep{Nixon2012}.  However, this criterion is difficult to satisfy in the long-run.
The relevant disk mass is the total amount of mass passing through it integrated
over time, and, as the binary's eccentricity approaches unity, $J_{\rm bin}$ falls.
Both effects make it easier to flip the binary around while still conserving
the total angular momentum of the system \citep{Roedig2014}. 
We have shown already that the angular momentum conveyed with mass
transfer is insufficient to accomplish this transition to a prograde
orientation before gravitational wave emission
brings the binary to merger; however, the situation changes when the
binary orbital plane and the disk plane are not exactly co-planar because
in that configuration
angular momentum can also be transferred between the two systems via
the mutual torque arising from the binary's quadrupole moment.

Suppose, then, that the angular momentum fed to the disk at large radius
is aligned not exactly opposite to the angular momentum of the binary, but only
approximately.   As shown by \cite{MK2013}, when the binary orbit is
circular, the inner portion of the disk aligns relatively quickly with the orbital
plane, but beyond a certain distance the disk maintains its intrinsic orientation.
Once the disk has reached a steady-state shape, the rate at which
angular momentum is delivered from the disk to the binary is $\sim \dot M j_T$, where
$j_T$ is the magnitude of the misaligned specific angular momentum in the disk at the
radius where the disk orientation transitions from aligned to misaligned, and the radius
of that transition is $R_T \sim (r/h)_T a$, where $(h/r)_T$ is the disk aspect ratio at $R_T$.

For the present application, this result must be generalized in two ways.   The first is
that their work assumed the binary has a circular orbit, so it must be extended to eccentric
orbits.  Following \cite{LLMech}, it is easy to show that the time-averaged
quadrupole moments of an eccentric binary are
\begin{subequations}
\begin{eqnarray}
Q_{xx} &=& \frac{M}{2}\frac{q}{(1+q)^2} a^2 (1 + 4e^2) \\
Q_{yy} &=& \frac{M}{2}\frac{q}{(1+q)^2} a^2 (1 - e^2)\, ,
\end{eqnarray}
\end{subequations}
where we define the $x-$ and $y-$
directions as parallel to the major and minor axes of the ellipse, respectively.
If the disk is tilted in the $\hat{e} - \hat{z}$ plane, where $\hat z \perp \hat x, \hat y$
and $\hat e = \cos\phi \hat x + \sin\phi \hat y$, the torque is proportional to
$Q_{xx} \cos^2\phi + Q_{yy} \sin^2\phi$. 

Because the quadrupolar torque is precessional, both the orientation of the disk tilt
plane $\phi$ and the misalignment angle $\theta$ can be functions of radius within the
disk.   The total binary-disk torque is then
\begin{multline}\label{eqn:torque}
{\cal T} = -\frac{3\pi}{2}\frac{q}{(1+q)^2} \Omega_T^2 a^2
\sin\theta_0 R_T^2 \Sigma_T \times \\
\int \, dx x^{-2}
\frac{\sin\theta(x)}{\sin\theta_0} \frac{\Sigma(x)}{\Sigma_T}
\left[(1 + 4e^2)\cos^2\phi(x) 
     + (1-e^2) \sin^2\phi(x)\right]\, ,
\end{multline}
where quantities with subscript $T$ are evaluated at $R_T$, beyond
which the disk's intrinsic alignment makes an angle $\theta_0$ with the binary.
The orbital frequency at this location
is $\Omega_T$, $\Sigma_T$ the surface density, and $x = r/R_T$. 
The form of this expression for ${\cal T}$ is written so as to make the integral both
dimensionless and order unity, as in \cite{MK2013}. Following their notation, we
designate it as ${\cal I}$, but now recognize that, unlike in the case
of circular binary orbits, it is a function of $e$ and precession
phase profile $\phi(r)$.

The second generalization is a consequence of the implicit assumption in \cite{MK2013}
that alignment of the binary remains incomplete even after the alignment front has propagated all the way
through the inner disk, so that the torque is concentrated near $r \simeq R_T$.  If, when
the system is first formed, the misaligned angular momentum contained in the disk within
a radius smaller than $R_T$ (the {\it steady-state} alignment transition radius) is
comparable to the angular momentum of the binary, the binary will have reoriented
before the alignment front arrives at the steady-state radius.

In fact, just this situation may often occur in the context of binary black holes.   Suppose
that the circumbinary disk extends from an inner radius $\simeq a$ (because it is obliquely
retrograde) out to an outer radius $r_{\rm out}$. As we did in the previous
section when discussing the approach to decoupling, we will describe the disk's surface
density profile in terms of a particular aspect ratio $h/r$, rather
than calculate the aspect ratio 
from a model for the disk's thermal state. The ratio between its misaligned
angular momentum and the binary's total angular momentum is then
\begin{equation}
\frac{J_{\rm disk}}{J_{\rm bin}} = 2\pi \sin\theta_0 \frac{\dot
  m}{\alpha \eta} (a/r_g)^{3/2} (r/h)^2 
\frac{ [(r_{\rm out}/a)^2 - 1] (1+q)^2}{q (1-e^2)^{1/2}} \frac{Gr_g}{\kappa_T c^2},
\end{equation}
where $\kappa_T$ is the Thomson opacity.   If the scale height varies as a function of radius,
$(r/h)^2$ should be regarded as the angular momentum-weighted mean of $(r/h)^2$.
Evaluated for our fiducial values, this ratio is
\begin{equation}\label{eq:angmomratio}
\frac{J_{\rm disk}}{J_{\rm bin}} \simeq 0.06\,  \sin\theta_0
\left(\frac{\alpha}{0.1}\right)^{-1} 
\frac{\dot m}{\eta} M_8 \left(\frac{a}{10^5r_g}\right)^{3/2} \left(\frac{r/h}{100}\right)^2  
  \frac{1}{(1 -e^2)^{1/2}} \frac{(1+q)^2}{q} \left[ (r_{\rm out}/a)^2 - 1\right].
\end{equation}
In other words,  for our fiducial values, the disk does not need to extend to a large multiple
of $a$ in order for its angular momentum content to match the binary's.  However,
it is also clear that this ratio is sensitive to a variety of parameters.

Suppose first that $J_{\rm disk} \gg J_{\rm bin}$.  In this case, as the alignment front
propagates out through the disk, the corresponding torque on the binary causes its
angular momentum to swing toward the direction of the disk's angular momentum.
Because the direction of the binary's angular momentum is initially nearly opposite to
that of the outer disk, this swing initially {\it increases} the angle between the two orbital
planes, the inclination angle growing exponentially if it is initially small
\citep{Scheuer1996}\footnote{\cite{Nixon2012} claimed otherwise, but he assumed a disk
without continuing accretion supply and ran for too short a time to test stability.}.

However, equation (\ref{eq:angmomratio}) suggests that there could also be disks of limited
radial extent containing less misaligned angular momentum than the binary, and the maximum $r_{\rm out}/a$
for which $J_{\rm disk} < J_{\rm bin}$ depends on the accretion rate, binary mass, etc.
When $r_{\rm out}$ is small enough for $J_{\rm disk} < J_{\rm bin}$, the disk counter-aligns
swiftly, i.e., it moves into the binary orbital plane, but orbiting
retrograde to the binary motion. 
Adapting the formalism of \cite{Sorathia2013} from the Lense-Thirring problem
they considered to the closely analogous problem of alignment with a binary, we find that the
alignment front moves outward at a rate
\begin{equation}
v_f(r) \simeq \frac{3}{8} \frac{q}{(1+q)^2} \frac{a^2 \Omega (r)}{r}
{\cal I}(r)\, .
\end{equation}
In this expression ${\cal I}$ is redefined so that it covers the region within $r$ rather
than within $R_T$.   Consequently $x$ in equation (\ref{eqn:torque})
becomes $r^\prime/r$ and $\Sigma_T$ becomes $\Sigma(r)$.
Thus, for order unity mass ratio in the binary, $v_f \sim (a/r)^2 v_{\rm orb}(r)$, not
much slower than the orbital speed, and the time for alignment to be achieved throughout 
a truncated disk is $\sim (r_{\rm out}/a)^2$ binary orbital periods at its outer
radius.

Once such a disk is entirely aligned, the binary exerts no further torque on the disk proper, but does
continue to interact with material at and beyond its outer boundary, $r_{\rm out}$.  Because
the torque decreases rapidly with increasing radius (the precession frequency is $\propto r^{-7/2}$),
only the nearest matter is significant.  Typically the mass just outside the disk should be
much smaller than the mass within the disk because the surface density of the disk is
$\sim 200 (\alpha/0.1)^{-1} (\dot m/\eta)[(r/h)/100]^2 (r/10a)^{-1/2}
(a/10^5 r_g)^{-1/2} \kappa_T^{-1}$, a surface density much greater than would
be expected from the surrounding interstellar medium.
Consequently, the torque that the binary continues to exert is on the material that has just
become attached to the disk, arriving with the intrinsic orientation of the mass supply: ${\cal T} = 
\sin \theta_0 \dot M r_{\rm out}^2 \Omega (r_{\rm out})$.  This torque
is smaller than the steady-state torque estimated through the formalism of \cite{MK2013} by
the ratio $(r_{\rm out}/R_T)^{1/2}$.

Whether the binary evolves to gravitational wave-driven merger faster
than it aligns with the external mass supply depends on whether $t_{\rm align}$, the time
required for this torque to change the orientation of the binary, is larger or smaller
than $t_{\rm trans}$, the time for accretion to drive binary  orbital evolution.   Because both
timescales depend on the angular momentum content of
the binary and the mass accretion rate, their ratio is very simple:
\begin{equation}
\frac{t_{\rm trans}}{t_{\rm align}} = \frac{1}{3} \sin \theta_0 (r_{\rm out}/a)^{1/2}.
\end{equation}
Thus, the maximum size of the circumbinary disk permitting retrograde orbital evolution to
reach its conclusion is limited by two constraints.  One is that the timescale for eccentricity growth must
be shorter than that for disk-binary alignment:
\begin{equation}
r_{\rm out} \lesssim \left(9/\sin^2 \theta_0\right) a.
\end{equation}
For moderate obliquity, $r_{\rm out}/a$ could be as large as $\simeq 10$--100 and still 
satisfy this condition. The other constraint is the one described above in equation (\ref{eq:angmomratio}),
namely that $J_{\rm disk}$ be $\lesssim J_{\rm bin}$.  This requirement
depends on more parameters. For our fiducial choices, it gives a
tighter constraint, $r_{\rm out} \lesssim 2a/\sqrt{\sin\theta_0}$, but this
constraint becomes similar to or looser than the timescale condition
when $(\dot m/\eta) M_8 (a/10^5 r_g)^{3/2}[(r/h)/100]^2  (1+q)^2/q
\lesssim 0.1$ (e.g., smaller accretion rates, more compact binaries,
or thicker outer disks).

It is worth pointing out that the timescale criterion we have used can be rephrased in
terms of the total angular momentum and mass budgets.   \citet{King2005} argued that
the binary can be flipped when $J_{\rm disk}/J_{\rm bin} > 2$.   This criterion applies
when there is no resupply of mass to the disk.   The alignment timescale in our case
should therefore be posed in terms of the ratio of the binary angular momentum to the
rate at which (misaligned) angular momentum is supplied to the
disk. The transition time $t_{\rm trans}$ is defined above in equation
(\ref{eqn:t_trans}), and is roughly the time it takes to accrete
$~20\%$ of the secondary's mass. 

Up to this point, we have ignored the question of whether such truncated disks might be
expected in Nature.   In fact, they may be reasonably common.   
Presumably what defines the outer edge of the circumbinary disk is the point at which
interstellar gas, following ballistic orbits in the gravitational potential created by
the binary black hole and the host galaxy, shocks against the disk's outer edge and
joins the accretion flow.  Another reason why the inventory of mass immediately outside
the disk should be much smaller than the disk mass is, in fact, that mass in this region travels
inward ballistically, whereas within the disk inward motion is dependent upon
the comparatively slow process of internal angular momentum transport.  The place where
external matter joins the disk could be at a radius as small
as that for which the specific angular momentum of the interstellar gas supports
a roughly circular orbit; alternatively, it could be at larger radius if the accretion flow (having this orientation)
has lasted long enough for the outer edge of the disk to be pushed outward significantly
by the outward angular momentum flux of accretion.   

Details of the formation of binary black holes are complex and depend upon numerous
parameters of the galaxy mergers preceding the formation of such a binary.  However, chaotic
motion of interstellar gas on the scale of parsecs to tens of parsecs
with orbital velocities of several hundred km/s can often be seen in
simulations of specific scenarios  \citep{Mayer2007,Roskar2014}. 
By comparison, the physical scale of the of the disks we have
been considering is similar or somewhat smaller, $\sim 5$~pc for our
fiducial parameters with $r_{\rm out}\sim 10a$. Thus, a truncated
obliquely retrograde disk of the sort we have discussed may well be
formed. 

\section{Discussion}

To form a circumbinary accretion disk, one requires a SMBH binary and a
supply of gas. Galactic mergers are almost certainly required to
provide the binary, but although mergers may enhance the availability of
gas, they are not strictly necessary, as we see plenty of examples of
AGN in non-merging galaxies.
Assuming a merger did take place and a SMBH binary has formed, 
if subsequent nuclear activity is driven by the stochastic
accretion of molecular clouds with random orientation relative to the
central SMBHB orbit, we expect it would be equally likely for
circumbinary accretion disks to form in the prograde or retrograde
orientations. 

With major mergers happening every $\sim 3-10$ Gyr
per galaxy in the local universe \citep{Lotz2011}, and typical
AGN lifetimes $\sim 10^{7}$ years \citep{Martini2004} with
duty cycle $\sim 10^{-2}$ \citep{Marconi2004}, we
should expect multiple accretion phases between mergers. For prograde
circumbinary disks, doubling the secondary mass leads to the binary orbit
shrinking by a factor of order unity \citep{Shi2012}, so many prograde
episodes would be necessary to reach the GW-dominated phase.
However, retrograde disks lead to much more rapid orbital evolution:
a mere 20\% increase in the mass of the secondary suffices. For the same
accretion rate at large radius, retrograde disks are more efficient in driving
binary orbital evolution than prograde disks not because they transmit more
mass to the binary [both cases maintain accretion flows close to inflow
equilibrium \citep{Farris2014,Shi2014}], but because retrograde
accretion changes the specific angular momentum of the binary by a
much larger factor.   The angular momentum carried to the binary by
accretion in the prograde case can in fact significantly counteract the angular
momentum lost by the binary through gravitational torques on the
circumbinary disk \citep{Shi2012}.

Thus, within this paradigm of multiple, stochastic accretion episodes
between mergers, there is a strong statistical bias toward such rapid
eccentricity evolution via accretion from retrograde disks that the GW-driven
state is achieved quickly.  When the mass accreted during
a single episode of constant angular momentum orientation is less than
roughly half the secondary mass, substantial orbital evolution occurs {\it only}
for roughly retrograde orientation.   Moreover, when substantial orbital
evolution does take place due to a retrograde accretion episode, it drives
the binary toward merger, so that no further episodes, retrograde or prograde,
occur. 

Torques couple binaries to obliquely-oriented circumbinary disks.    When
$\vec J_{\rm disk} \cdot \vec J_{\rm binary} < 0$ and enough mass is available
(either in the initial disk or over the course of its feeding) that the total disk
angular momentum is larger than the binary's, these torques can
``flip" the binary orbit so that it aligns with the disk plane and orbits in the
same sense \citep{Scheuer1996,King2005}.   However, we have shown
that if the disk is fed at radii not too much larger than
the binary semimajor axis, the angular momentum instantaneously stored in the disk can be small
enough, and the evolution of the binary orbit caused by retrograde accretion can
be rapid enough, that the binary is driven to the gravitational wave-dominated stage {\it before}
its orbit is flipped.    Thus, the condition for retrograde accretion to cause
merger before the binary orbit is reoriented amounts to a condition on the
disk's outer radius relative to the scale of the binary orbit.

As a result, high-eccentricity and ultimately GW-driven SMBHBs may be a common
result of gas-driven black hole mergers unless any of three
conditions applies: accretion episodes frequently occur in which the
total mass accreted is large enough to cause substantial binary
orbital evolution even if the rotation sense is prograde;  the
orientation of the accreted matter's angular momentum is consistently correlated with that of the
binary black hole system; or the outer edge of the circumbinary disk is so far outside
the binary that retrograde systems are flipped before their orbits
evolve.   In evaluating this statement, we emphasize the 
qualifier ``gas-driven" because gas accretion may not always be the principal element in
binary black hole evolution: stellar interactions may be more important in some
cases. This is especially relevant for the massive
  gas-poor elliptical galaxies that host the most massive black holes,
  and which are thought to dominate the PTA signal. These galaxies
  also exhibit decreased nuclear activity.
It is also possible that even when stellar interactions are weak, the
evolution of the circumbinary disk at large radii may be complicated
by self-gravity and internal star formation.

One consequence of the gas-driven, rapid merger scenario would be a distinct
absence of binary quasars with separations between $\sim 0.01-1$ pc, a
source population that has been notoriously difficult to
detect \citep{Eracleous2012}.  To the degree that gas-driven accretion
dominates binary evolution in PTA sources, the effects we have pointed
out would have profound effects on the potentially-observable GW
signal, effectively moving most of the GW power from 
the nHz to the $\mu$Hz band \citep{Sesana2013}. By the time such systems reach the mHz
band, they should be highly circularized. 

Because the GW evolution time drops so rapidly with increasing
eccentricity, it is likely that at some point before black hole
merger,  the circumbinary disk will no longer be able to stretch
inward as rapidly as the binary apocenter distance shrinks; that is, 
the the system will ``decouple.'' Yet even in such a situation,
there still might be enough low-density, highly magnetized gas in the
central gap to lead to an observable EM signal. Understanding the
precise nature of such a system will require new 3D global MHD
simulations that can sufficiently resolve both the disk and the gas
flow onto the individual black holes over long periods of time. 

\section*{Acknowledgments}

This work was partially supported by National Science Foundation grant
AST-1028111 and NASA grant ATP12-0139. We thank Cole Miller for
helpful comments and discussion.    We also thank the referee for leading
us to pursue the properties of binary--circumbinary disk alignment much
further than we had initially.

\bibliography{circumbin.bib}

\end{document}